# Rortex and comparison with eigenvalue-based vortex identification criteria

Yisheng Gao[1] and Chaoqun Liu[1,a)]

[1]*Department of Mathematics, University of Texas at Arlington, Arlington, Texas 76019, USA*

Most of the current Eulerian vortex identification criteria, including the Q criterion and the $\lambda_{ci}$ criterion, are exclusively determined by the eigenvalues of the velocity gradient tensor or the related invariants and thereby can be regarded as eigenvalue-based criteria. However, these criteria will be plagued with two shortcomings: (1) these criteria fail to identify the swirl axis or orientation; (2) these criteria are prone to severe contamination by shearing. To address these issues, a new vector named Rortex which represents the local fluid rotation was proposed in our previous work. In this paper, an alternative eigenvector-based definition of Rortex is introduced. The direction of Rortex, which represents the possible axis of the local rotation, is determined by the real eigenvector of the velocity gradient tensor. And then the rotational strength obtained in the plane perpendicular to the possible axis is used to define the magnitude of Rortex. This new equivalent definition allows a much more efficient implementation. Furthermore, a systematic interpretation of scalar, vector and tensor versions of Rortex is presented. By relying on the tensor interpretation, the velocity gradient tensor is decomposed to a rigid rotation part and a non-rotational part including shearing, stretching and compression, different from the traditional symmetric and anti-symmetric tensor decomposition. It can be observed that shearing always manifests its effect on the imaginary part of the complex eigenvalues and consequently contaminates eigenvalue-based criteria, while Rortex can exclude the shearing contamination and accurately quantify the local rotational strength. In addition, in contrast to eigenvalue-based criteria, not only the iso-surface of Rortex, but also the Rortex vectors and the Rortex lines can be applied to investigate vortical structures. Several comparative

a)Email: cliu@uta.edu

studies on simple examples and realistic flows are studied to confirm the superiority of Rortex.

## I. INTRODUCTION

Vortical structures, also referred to as coherent turbulent structures,[1-4] are generally acknowledged as one of the most salient characteristics of turbulent flows and occupy a pivotal role in turbulence generation and sustenance since a conceptual model of hairpin vortex was proposed by Theodorsen.[5] Several important coherent structures have been identified, including vortex "worms" in isotropic turbulence,[6,7] hairpin vortices in wall-bounded turbulence,[8-10] quasi-streamwise vortices[4,11,12] and vortex braids in turbulent shear layers,[13,14] etc. Naturally, the ubiquity and significance of such spatially coherent, temporally evolving vortical motions in transitional and turbulent flows necessitate an unambiguous and rigorous definition of vortex for the comprehensive and thorough investigation of these sophisticated phenomenon. Unfortunately, although vortex can be intuitively recognized as the rotational/swirling motion of fluids and has been intensively studied for more than one hundred years, a sound and universally-accepted definition of vortex is yet to be achieved in fluid mechanics,[15,16] which is possibly one of the chief obstacles causing considerable confusions in visualizing and understanding vortical structures.[17-19]

The classic definition of vortex is associated with vorticity which possesses a clear mathematic definition, namely the curl of the velocity vector. As early as 1858, Helmholtz first considered a vorticity tube with infinitesimal cross-section as a vortex filament,[20] which was followed by Lamb to simply call a vortex filament as a vortex in his classic monograph.[21] Similarly, Nitsche declares that "A vortex is commonly associated with the rotational motion of fluid around a common centerline. It is defined by the vorticity in the fluid, which measures the rate of local fluid rotation."[22] And several contemporary treatises on vortex dynamics (vorticity dynamics) advocate vorticity-based definitions as well. For example, Saffman's book defines a vortex as a "finite volume of vorticity immersed in irrotational fluid."[23] Wu et al. suggest that "a vortex is a connected region with high concentration of vorticity compared with its surrounding".[15] Since vorticity is well-defined, vorticity dynamics has been systematically developed for the generation and evolution of vorticity and applied in the study of vortical-flow stability and vortical structures in transitional and turbulent flows.[15,23,24] However, the use of vorticity will run into severe difficulties in viscous flows, especially in turbulence, because vorticity is unable to distinguish between a real vortical region and a shear layer region. It is not uncommon that the average shear force generated by the no-slip wall is so strong in the boundary layer of a laminar flow plane that an extremely large amount of vorticity exists but no vortical motions will be observed in some near-wall regions.[25] Jeong and Hussain provide a discussion on the inadequacy of iso-vorticity surfaces for detecting vortices.[26] Meanwhile, the determination of an appropriate threshold above which vorticity can be considered as high concentrated is a common problem in practice.[27] On the other hand, it has been noticed by several researchers that the local vorticity vector is not always aligned with the direction of vortical structures in turbulent wall-bounded flows, especially at locations close to the wall. Gao et al. analyze the vortex populations in turbulent wall-bounded flows to demonstrate the vorticity can be somewhat misaligned with the vortex core direction.[28] Zhou et al. find that the vorticity vector angle is significantly larger than the local inclination of the vortical structure over almost the entire length of the quasi-streamwise vortex in the channel flow.[29] Pirozzoli et al. also show the differences between the local vorticity direction and the vortex core orientation in a supersonic turbulent boundary layer.[30] Furthermore, the maximum vorticity does not necessarily occur in the central region of vortical structures. As pointed out by Robinson, "the association between regions of strong vorticity and actual vortices can be rather weak in the turbulent boundary layer, especially in the near wall region."[31] Wang et al. obtain a similar result that the magnitude of vorticity can be substantially reduced along vorticity lines entering the vortex core region near the solid wall in a flat plate boundary layer.[25]

The problems of vorticity for the identification and visualization of vortical structures in turbulence motivate the rapid development of vortex identification methods, including intuitive measures, Eulerian velocity-gradient-based criteria and Lagrangian objective criteria, etc. The common intuitive indicators, such as local pressure minima and closed or spiraling streamlines and pathlines, though seemingly obvious and easy to understand, suffer from serious troubles in identifying vortices, which is explained in great detail by Jeong and Hussain.[26] Most of the currently popular Eulerian vortex identification criteria are based on the analysis of the velocity gradient tensor. More specifically, these criteria are exclusively determined by the eigenvalues of the velocity gradient tensor or the related invariants and thereby can be regarded as eigenvalue-based criteria.[32] For example, the Q criterion defines vortices as the region with positive second invariant of the velocity gradient tensor.[33] The Δ criterion employs the discriminant of the characteristic equation to identify the region where the velocity gradient tensor has complex eigenvalues.[34,35,16] The $\lambda_{ci}$ criterion uses the (positive) imaginary part of the complex eigenvalue to determine the swirling strength.[29] And the $\lambda_2$ criterion is based on the second-largest eigenvalue of $\boldsymbol{S}^2 + \boldsymbol{\Omega}^2$ ($\boldsymbol{S}$ and $\boldsymbol{\Omega}$ represent the symmetric and the antisymmetric parts of the velocity gradient tensor, respectively). Note that, generally, $\lambda_2$ cannot be expressed in terms of the eigenvalues of the velocity gradient tensor. However, in the special case when the eigenvectors are orthonormal, $\lambda_2$ can be exclusively determined by the eigenvalues.[26] One remarkable feature of these criteria is Galilean invariant, since these criteria are based on the kinematics implied by the velocity gradient tensor which is the same in all inertial frames. Another cardinal virtue is that these methods are concerned with identifying vortex cores, and thus can discriminate against shear layers, offering more detectable vortical structures. Usually, these criteria require user-specified thresholds. It is vital to determine an appropriate threshold, since different thresholds will indicate different vortical structures. For instance, even if the same DNS data on the late boundary layer transition are examined, "vortex breakdown" will be exposed with the use of a large threshold for the $\lambda_2$ criterion while no "vortex breakdown" will be observed with a small threshold.[36] Accordingly, the educed structures obtained from these criteria should be interpreted with care. The choice of thresholds has been studied by Cucitore at el,[37] Chakraborty et al.,[32] and del Álamo at el.[38] As a remedy, relative values can be employed to avoid the usage of case-related thresholds, and one such example is the Ω method proposed by Liu et al.[39] Despite of the widespread use, these eigenvalue-based criteria are not always satisfactory. One obvious drawback is the inadequacy of identifying the swirl axis or orientation. Since the vortex is recognized as the rotational motion of fluids, it is expected that the swirl axis or orientation will provide information for the analysis of vortical structures. Nonetheless, the existing eigenvalue-based criteria are scalar-valued criteria and thus unable to identify the swirl axis or orientation. Another shortcoming is the contamination by shearing. Recently, the $\lambda_{ci}$ criterion has been found to be serious contaminated by shearing motion.[40,41] In fact, as described below, other eigenvalue-based criteria will suffer from the same problem, as long as the criterion is associated with the complex eigenvalues. This issue prompts Kolář to formulate a triple decomposition from which the residual vorticity can be obtained after the extraction of an effective pure shearing motion and represents a direct and accurate measure of the pure rigid-body rotation of a fluid element.[42] However, the triple decomposition requires a basic reference frame to be first determined. Searching for the basic reference frame in 3D cases will result in an expensive optimization problem for every point in the flow field, which limits the applicability of the method. And the triple decomposition has not yet been thoroughly investigated for 3D cases. Hence, Kolář et al. introduce the concept of the average corotation of line segments near a point to reduce the computational overhead.[43] In addition to the widely used Eulerian vortex identification methods, some objective Lagrangian vortex identification methods have been developed to study the vortex structures involved in the rotating reference frame.[3,44] For extensive overview of the currently available vortex identification methods, one can refer to review papers by Epps[27] and Zhang et al.[45]

To address the above-mentioned issues of the existing eigenvalue-based criteria, a new vector quantity, which is called vortex vector or Rortex, was proposed and investigated in our previous works.[46,47] In this paper, an alternative eigenvector-based definition of Rortex is presented. The direction of Rortex, which represents the possible axis of the local rotation, is determined by the real eigenvector of the velocity gradient tensor. And then the rotational strength determined in the plane perpendicular to the possible axis is used to define the magnitude of Rortex. The rotational strength of Rortex is equivalent to Kolář's residual vorticity in 2D cases, but Kolář's triple decomposition has yet to be fully studied in 3D cases and thus the result is unclear. The main distinguishing feature of Rortex is that Rortex is eigenvector-based and the magnitude (rotational strength) is strongly relevant to the direction of the real eigenvector. Although Gao et al. use the real eigenvector to indicate the orientation about which the flow swirls, they choose the imaginary part of the complex eigenvalues as the swirling strength and the swirling strength is determined independent of the choice of the orientation.[28] The present eigenvector-based definition is mathematically equivalent to our previous one but significantly improves the computational efficiency. Furthermore, a complete and systematic interpretation of scalar, vector and tensor versions of Rortex is presented to provide a unified and clear characterization of the local fluid rotation. The scalar represents the local rotational strength, the vector offers the local swirl axis and the tensor extracts the local rigidly rotational part of the velocity gradient tenor. Especially, the tensor interpretation brings a new decomposition of the velocity gradient tensor to investigate the analytical relations between Rortex and eigenvalue-based criteria. The velocity gradient tensor in a special reference frame is examined to indicate that shearing always manifests its effect on the imaginary part of the complex eigenvalues and consequently contaminates eigenvalue-based criteria. In contrast, Rortex can exclude the shearing contamination and accurately quantify the local rotational strength. A comprehensive comparison of Rortex with the Q criterion and the $\lambda_{ci}$ criterion on several simple examples and realistic flows is carried out to confirm the superiority of Rortex.

The remainder of the paper is organized as follows. In Section II, our previous definition of Rortex is revisited, followed by an eigenvector-based definition, and the new implementation is also provided. The systematic interpretation of scalar, vector and tensor versions of Rortex and the analytical comparison of Rortex and eigenvalue-based criteria are elaborated in Section III. Several comparative studies on simple examples are carried out in Section IV. Section V shows the comparison of Rortex and eigenvalue-based criteria on the DNS data of the boundary layer transition over a flat plate. The conclusions are summarized in the last section.

## II. EIGENVECTOR-BASED DEFINITION OF RORTEX

### A. Four principles

To reasonably define a vortex vector or Rortex, we propose the following principles:

(1) **Local**. Although a vortex is regarded as a non-local flow motion, the presence of viscosity in real flows leads to the continuity of the kinematic features of the flow field[32] and numerous studies have suggested that the cores of vortical structures in turbulent flows are well localized in space.[26] Moreover, critical-point concepts based on local kinematics of the flow field have successfully provided a general description of 3D steady and unsteady flow pattern.[16,48] And non-locality commonly implies much more complexity in computation.

(2) **Galilean Invariant**. It means that the definition is the same in all inertial frames. This principle is followed by many Eulerian vortex identification criteria.[26,32,33] Objectivity may be preferred when involved in a more general motion of the reference frame,[49,50] but it is beyond the subject of the present study.

(3) **Unique**. The description of the local rigidly rotation must be accurate and unique. It requires the exclusion of the contamination by shearing.

(4) **Systematical**. The definition will contain a scalar version which is the strength of the rigid rotation, a vector version which provides both the rotation axis and rotation strength, and a tensor version which represents the rigid rotation part of the velocity gradient tensor.

## B. Previous definition of Rortex

Based on the above principles, the concept of the local fluid rotation and a vector named vortex vector or Rortex which represents the local fluid rotation are introduced in our previous work.[46,47] The direction of Rortex is determined by the direction of the local rotation axis Z, and the magnitude of Rortex is defined by the rotational strength of the local fluid rotation, which is determined in the XY plane perpendicular to the Z-axis. If U, V and W are velocity components along the X, Y and Z axes respectively, the matrix representation of the velocity gradient tensor in the XYZ-frame can be written as

$$\nabla \vec{V} = \begin{bmatrix} \frac{\partial U}{\partial X} & \frac{\partial U}{\partial Y} & 0 \\ \frac{\partial V}{\partial X} & \frac{\partial V}{\partial Y} & 0 \\ \frac{\partial W}{\partial X} & \frac{\partial W}{\partial Y} & \frac{\partial W}{\partial Z} \end{bmatrix} \qquad (1)$$

Generally, the z-axis in the original $xyz$-frame is not parallel to the Z-axis, so the velocity gradient tensor in the origin $xyz$-frame

$$\nabla \vec{v} = \begin{bmatrix} \frac{\partial u}{\partial x} & \frac{\partial u}{\partial y} & \frac{\partial u}{\partial z} \\ \frac{\partial v}{\partial x} & \frac{\partial v}{\partial y} & \frac{\partial v}{\partial z} \\ \frac{\partial w}{\partial x} & \frac{\partial w}{\partial y} & \frac{\partial w}{\partial z} \end{bmatrix} \quad (2)$$

cannot fulfill Eq. (1). Thus, a coordinate transformation is required to rotate the z-axis to the Z-axis. There exists a corresponding transformation between $\nabla \vec{V}$ and $\nabla \vec{v}$

$$\nabla \vec{V} = \boldsymbol{Q} \nabla \vec{v} \boldsymbol{Q}^{-1} \quad (3)$$

where $\boldsymbol{Q}$ is a rotation matrix and

$$\boldsymbol{Q}^{-1} = \boldsymbol{Q}^{\mathrm{T}} \quad (4)$$

In Ref. 47, the existence of the local rotation axis Z was proved through real Schur decomposition.[52] The direction of the local rotation axis Z can be obtained by solving a nonlinear system of equations through the Newton-iterative method[46] or by a fast algorithm based on real Schur decomposition.[47]

If the direction of the Z-axis in the $xyz$-frame is given by $\vec{r} = [r_x, r_y, r_z]^T$,

$$\vec{r} = \boldsymbol{Q}^T \begin{bmatrix} 0 \\ 0 \\ 1 \end{bmatrix} \quad (5)$$

and

$$\boldsymbol{Q}\vec{r} = \begin{bmatrix} 0 \\ 0 \\ 1 \end{bmatrix} \quad (6)$$

represents the direction of the local rotation axis Z in the XYZ-frame.

Once the local rotation axis Z is obtained, the rotation strength is determined in the XY plane perpendicular to the local rotation axis Z. This can be achieved by a second coordinate rotation in the XY plane. When the XYZ-frame is rotated around the Z-axis by an angle θ, the velocity gradient tensor will become

$$\nabla \vec{V}_\theta = \boldsymbol{P} \nabla \vec{V} \boldsymbol{P}^{-1} \quad (7)$$

where **P** is the rotation matrix around the Z-axis and can be written as

$$P = \begin{bmatrix} \cos\theta & \sin\theta & 0 \\ -\sin\theta & \cos\theta & 0 \\ 0 & 0 & 1 \end{bmatrix} \tag{8}$$

$$P^{-1} = P^T = \begin{bmatrix} \cos\theta & -\sin\theta & 0 \\ \sin\theta & \cos\theta & 0 \\ 0 & 0 & 1 \end{bmatrix} \tag{9}$$

So, we have

$$\frac{\partial U}{\partial Y}\Big|_\theta = \alpha \sin(2\theta + \varphi) - \beta \tag{10a}$$

$$\frac{\partial V}{\partial X}\Big|_\theta = \alpha \sin(2\theta + \varphi) + \beta \tag{10b}$$

$$\frac{\partial U}{\partial X}\Big|_\theta = \alpha \cos(2\theta + \varphi) + \frac{1}{2}\left(\frac{\partial U}{\partial X} + \frac{\partial V}{\partial Y}\right) \tag{10c}$$

$$\frac{\partial V}{\partial Y}\Big|_\theta = -\alpha \cos(2\theta + \varphi) + \frac{1}{2}\left(\frac{\partial U}{\partial X} + \frac{\partial V}{\partial Y}\right) \tag{10d}$$

where

$$\alpha = \frac{1}{2}\sqrt{\left(\frac{\partial V}{\partial Y} - \frac{\partial U}{\partial X}\right)^2 + \left(\frac{\partial V}{\partial X} + \frac{\partial U}{\partial Y}\right)^2} \tag{11}$$

$$\beta = \frac{1}{2}\left(\frac{\partial V}{\partial X} - \frac{\partial U}{\partial Y}\right) \tag{12}$$

$$\varphi = \begin{cases} \arctan\left(\frac{\frac{\partial V}{\partial X} + \frac{\partial U}{\partial Y}}{\frac{\partial V}{\partial Y} - \frac{\partial U}{\partial X}}\right), & \frac{\partial V}{\partial Y} - \frac{\partial U}{\partial X} \neq 0 \\ \frac{\pi}{2}, & \frac{\partial V}{\partial Y} - \frac{\partial U}{\partial X} = 0, \frac{\partial V}{\partial X} + \frac{\partial U}{\partial Y} > 0 \\ -\frac{\pi}{2}, & \frac{\partial V}{\partial Y} - \frac{\partial U}{\partial X} = 0, \frac{\partial V}{\partial X} + \frac{\partial U}{\partial Y} < 0 \end{cases} \tag{13}$$

(Note: If $\frac{\partial V}{\partial Y} - \frac{\partial U}{\partial X} = 0, \frac{\partial V}{\partial X} + \frac{\partial U}{\partial Y} = 0, \frac{\partial V}{\partial X} = \beta, \frac{\partial U}{\partial Y} = -\beta$ for any $\theta$, thus $\varphi$ is not needed.)

The criterion to determine the existence of local fluid rotation in the XY plane is

$$g_{Zmin} = \beta^2 - \alpha^2 > 0 \tag{14}$$

And $\frac{\partial U}{\partial Y}$ can be regarded as the angular velocity of the fluid at the azimuth angle θ relative to the point

$$\omega_\theta = \frac{\partial U}{\partial Y}\Big|_\theta = \alpha \sin(2\theta + \varphi) - \beta \tag{15}$$

Since $\omega_\theta$ will change with the change of the azimuth angle θ, the fluid-rotational angular velocity in the XY plane is defined as the absolute minimum of Eq. (15)

$$\omega_{rot} = \omega_{\theta min} = \begin{cases} -(\beta + \alpha), & g_{Zmin} > 0 \\ 0, & g_{Zmin} \leq 0 \end{cases} \quad (16)$$

Here, we assume $\beta < 0$. If $\beta > 0$, we can rotate the local rotation axis to the opposite direction to make $\beta$ positive. The local rotation strength (the magnitude of Rortex) is defined as twice the fluid-rotational angular velocity

$$R = \begin{cases} -2(\beta + \alpha), & g_{Zmin} > 0 \\ 0, & g_{Zmin} \leq 0 \end{cases} \quad (17)$$

The factor 2 is related to using 1/2 in the expression for the 2-D vorticity tensor component. It should be noted that Eq. (17) is equivalent to Kolář's residual vorticity in 2D cases.[42,43] But Kolář's triple decomposition is yet to thoroughly examined for general 3D cases and our numerical tests indicate that the rotation axis and the rotational strength of Rortex are totally different from Kolář's results in 3D cases.

## C. Eigenvector-based definition of Rortex

Our previous work provides a physical description of Rortex, but the relation between Rortex and the eigenvalues of the velocity gradient tensor and eigenvalue-based criteria is unclear. It motivates the present study.

**Definition 1**: A local rotation axis is defined as the direction of $\vec{r}$ where $d\vec{v} = \alpha d\vec{r}$.

This definition means that in the direction of the local rotation axis, there is no cross-velocity gradient. For example, if the z-axis is the rotation axis in a reference frame, the velocity can only increase or decrease along the z-axis, which means only $dw \neq 0$, but $du = 0$ and $dv = 0$. Accordingly, we can obtain the following theorem:

**Theorem 1.** The direction of the local rotation axis is the real eigenvector of the velocity gradient tensor $\nabla \vec{v}$.

Proof: If $\vec{r} = [r_x, r_y, r_z]^T$ represents the direction of the local rotation axis, we have $d\vec{v} = \alpha d\vec{r}$. On the other hand, from the definition of the velocity gradient tensor

$$d\vec{v} = \nabla \vec{v} \cdot d\vec{r} \quad (18)$$

Therefore,

$$\nabla \vec{v} \cdot d\vec{r} = \alpha d\vec{r} \tag{19}$$

and

$$\nabla \vec{v} \cdot \vec{r} = \alpha \vec{r} \tag{20}$$

which means $\vec{r}$ is the real eigenvector of $\nabla \vec{v}$ and $\alpha$ is the real eigenvalue.

The alternative definition of the direction of Rortex is equivalent to our previous one. If $\vec{r}$ is the real eigenvector of the velocity gradient tensor $\nabla \vec{v}$, then we have

$$\nabla \vec{v} \cdot \vec{r} = \alpha \vec{r} \tag{21}$$

Under the coordinate rotation $\boldsymbol{Q}$ which rotates the z-axis to the direction of $\vec{r}$, it can be written as

$$\boldsymbol{Q}\nabla\vec{v}\boldsymbol{Q}^{\mathrm{T}}\boldsymbol{Q}\vec{r} = \alpha \boldsymbol{Q}\vec{r} \tag{22}$$

According to Eq. (3), we can find

$$\nabla \vec{V} \cdot \boldsymbol{Q}\vec{r} = \alpha \boldsymbol{Q}\vec{r} \tag{23}$$

According to Eq. (6), $\boldsymbol{Q}\vec{r}$ is the direction of the local rotation in the XYZ-frame.

Conversely, the vector $\begin{bmatrix} 0 \\ 0 \\ 1 \end{bmatrix}$ is the real eigenvector of $\nabla \vec{V}$ in the XYZ-frame, since

$$\nabla \vec{V} \cdot \begin{bmatrix} 0 \\ 0 \\ 1 \end{bmatrix} = \begin{bmatrix} \frac{\partial U}{\partial X} & \frac{\partial U}{\partial Y} & 0 \\ \frac{\partial V}{\partial X} & \frac{\partial V}{\partial Y} & 0 \\ \frac{\partial W}{\partial X} & \frac{\partial W}{\partial Y} & \frac{\partial W}{\partial Z} \end{bmatrix} \begin{bmatrix} 0 \\ 0 \\ 1 \end{bmatrix} = \frac{\partial W}{\partial Z} \begin{bmatrix} 0 \\ 0 \\ 1 \end{bmatrix} \tag{24}$$

If we rotate the XYZ frame back to the origin $xyz$-frame, we have

$$\boldsymbol{Q}^{\mathrm{T}}\nabla\vec{V}\boldsymbol{Q}\boldsymbol{Q}^{\mathrm{T}}\begin{bmatrix} 0 \\ 0 \\ 1 \end{bmatrix} = \boldsymbol{Q}^{\mathrm{T}}\frac{\partial W}{\partial Z}\begin{bmatrix} 0 \\ 0 \\ 1 \end{bmatrix} \tag{25}$$

$$\nabla \vec{v} \cdot \left(\boldsymbol{Q}^{\mathrm{T}}\begin{bmatrix} 0 \\ 0 \\ 1 \end{bmatrix}\right) = \frac{\partial W}{\partial Z}\left(\boldsymbol{Q}^{\mathrm{T}}\begin{bmatrix} 0 \\ 0 \\ 1 \end{bmatrix}\right) \tag{26}$$

Through Eq. (5), $\vec{r} = \boldsymbol{Q}^{\mathrm{T}}\begin{bmatrix} 0 \\ 0 \\ 1 \end{bmatrix}$ represents the real eigenvector of the velocity gradient tensor $\nabla \vec{v}$ in the $xyz$-frame.

The definition of the rotational strength is the same as the previous one. It is determined in the plane perpendicular to the direction of the real eigenvector by Eq. (17).

It should be noted that when the velocity gradient tensor has three real eigenvalues, there exist more than one real eigenvectors, which implies the existence of multiple possible axes. However, according to our previous work,[47] when there exist three real eigenvalues $\lambda_1$, $\lambda_2$ and $\lambda_3$, $\nabla \vec{V}$ will become a lower triangular matrix which can be written as

$$\nabla \vec{V} = \begin{bmatrix} \lambda_1 & 0 & 0 \\ \frac{\partial V}{\partial X} & \lambda_2 & 0 \\ \frac{\partial W}{\partial X} & \frac{\partial W}{\partial Y} & \lambda_3 \end{bmatrix} \tag{27}$$

In this case, we have

$$\alpha = \frac{1}{2}\sqrt{(\lambda_2 - \lambda_1)^2 + \left(\frac{\partial V}{\partial X}\right)^2} \tag{28}$$

$$\beta = \frac{1}{2}\left(\frac{\partial V}{\partial X}\right) \tag{29}$$

Since $\alpha \geq \beta$, the rotation strength R given by Eq. (17) will be equal to zero. Therefore, Rortex is a zero vector in this case, which is consistent to our definition. Non-zero Rortex exists only if the velocity gradient tensor has one real eigenvalue and two complex eigenvalues. So, Rortex is equivalent to the $\Delta$ criterion and the $\lambda_{ci}$ criterion when a zero threshold is applied.

In Ref. 47, we use real Schur decomposition to prove the existence of the (possible) local rotation axis. But the uniqueness is not mentioned. Through the above eigenvector-based definition, the existence and uniqueness of the (possible) local rotation axis can be immediately proved from the existence and uniqueness (up to sign) of the normalized real eigenvector of the velocity gradient tensor when there exist a pair of complex eigenvalues.

**D. Calculation procedure for Rortex**

By relying on the eigenvector-based definition, the use the Newton-iterative method or real Schur decomposition, applied in our previous work,[46,47] can be avoided, making a significantly simplified implementation. The complete calculation procedure consists of the following steps:

1) Compute the velocity gradient tensor $\nabla \vec{v}$ in the $xyz$-frame;

2) Calculate the real eigenvalue $\lambda_r$ of the velocity gradient tensor $\nabla \vec{v}$ when the complex eigenvalues exists (the analytical expression is provided in Appendix A);

3) Calculate the (normalized) real eigenvector $\vec{r} = [r_x, r_y, r_z]^T$ corresponding to the real eigenvalue $\lambda_r$ (the analytical expression is provided in Appendix B);

4) Calculate the rotation matrix $Q^*$ using Rodrigues' rotation formula;[53]

$$Q^* = \begin{bmatrix} \cos\phi + r_x^2(1-\cos\phi) & r_x r_y (1-\cos\phi) - r_z \sin\phi & r_x r_z (1-\cos\phi) + r_y \sin\phi \\ r_y r_x (1-\cos\phi) + r_z \sin\phi & \cos\phi + r_y^2(1-\cos\phi) & r_y r_z (1-\cos\phi) - r_x \sin\phi \\ r_z r_x (1-\cos\phi) - r_y \sin\phi & r_z r_y (1-\cos\phi) + r_x \sin\phi & \cos\phi + r_z^2(1-\cos\phi) \end{bmatrix} \quad (30)$$

$$\phi = \mathrm{acos}(c) \quad (31)$$

$$c = \begin{bmatrix} 0 \\ 0 \\ 1 \end{bmatrix} \cdot \vec{r} \quad (32)$$

5) Obtain the velocity gradient tensor $\nabla \vec{V}$ in the XYZ frame via

$$\nabla \vec{V} = Q^{*T} \nabla \vec{v} Q^* \quad (33)$$

6) Calculate $\alpha$ and $\beta$ using Eqs. (11) and (12);

7) Obtain $R$ according to the signs of $\alpha^2 - \beta^2$ (Here, we assume $\beta < 0$. If not the case, we can rotate the local rotation axis $\vec{r}$ to the opposite direction to make $\beta$ positive. In addition, $R$ is invariant in the XY plane, so the calculation of the rotation matrix $P$ can be avoided.)

$$R = \begin{cases} -2(\beta + \alpha), & \alpha^2 - \beta^2 < 0 \\ 0, & \alpha^2 - \beta^2 \geq 0 \end{cases} \quad (34)$$

8) Compute Rortex $\vec{R}$ via

$$\vec{R} = R\vec{r} \quad (35)$$

The eigenvector-based definition brings remarkably improvement to the computational efficiency. In our earliest implementation, the direction of Rortex was obtained by solving a nonlinear system of equations through the Newton-iterative method.[46] In Ref. 47, a fast algorithm based on real Schur decomposition was proposed to reduce the computational cost. The real Schur decomposition is performed using a standard numerical linear algebra library LAPACK.[54] Table 1 illustrates the calculation time of our previous and present methods for the DNS data consisting of about 60 million points. The calculation time of the $\lambda_{ci}$ criterion is presented as well. All the calculations are run on a MacBook Pro (Late 2013) laptop with 2.0 GHz CPU and 8GB memory. It can be observed that the calculation time of the present method is reduced by one order of magnitude compared to our previous methods and comparable to that of the $\lambda_{ci}$ criterion.

TABLE 1. The calculation times of different methods for the DNS data.

| Method | Newton-iterative | Real Schur decomposition | $\lambda_{ci}$ | Present |
|---|---|---|---|---|
| Time (second) | 264 | 120 | 7.5 | 11.3 |

**E. Systematic interpretation of scalar, vector and tensor versions of Rortex and velocity gradient tensor decomposition**

Although Rortex is defined as a vector, we can propose a tensor interpretation of Rortex. When the absolute minimum of Eq. (15) is achieved, we have $2\theta + \varphi = -\pi/2$ (assume $\beta < 0$). Hence, the velocity gradient tensor given by Eq. (7) becomes

$$\nabla \vec{V}_{\theta min} = \begin{bmatrix} \frac{1}{2}(\frac{\partial U}{\partial X} + \frac{\partial V}{\partial Y}) & -(\beta + \alpha) & 0 \\ \beta - \alpha & \frac{1}{2}(\frac{\partial U}{\partial X} + \frac{\partial V}{\partial Y}) & 0 \\ \frac{\partial W}{\partial X}\big|_{\theta min} & \frac{\partial W}{\partial Y}\big|_{\theta min} & \frac{\partial W}{\partial Z}\big|_{\theta min} \end{bmatrix} \quad (36)$$

If $\lambda_{cr}$ represents the real part of the complex eigenvalues, $\lambda_{ci}$ the imaginary part of the complex eigenvalues and $\lambda_r$ the real eigenvalue, we can obtain

$$\lambda_r = \frac{\partial W}{\partial Z}\bigg|_{\theta min} \quad (37)$$

$$\lambda_{cr} = \frac{1}{2}(\frac{\partial U}{\partial X} + \frac{\partial V}{\partial Y}) \quad (38)$$

$$\lambda_{ci} = \sqrt{\beta^2 - \alpha^2} \quad (39)$$

Eq. (36) can be decomposed into two parts

$$\nabla \vec{V}_{\theta min} = \begin{bmatrix} \lambda_{cr} & \phi & 0 \\ -(\phi+\varepsilon) & \lambda_{cr} & 0 \\ \xi & \eta & \lambda_r \end{bmatrix} = \boldsymbol{R}+\boldsymbol{S} \tag{40}$$

$$\boldsymbol{R} = \begin{bmatrix} 0 & \phi & 0 \\ -\phi & 0 & 0 \\ 0 & 0 & 0 \end{bmatrix} \tag{41}$$

$$\boldsymbol{S} = \begin{bmatrix} \lambda_{cr} & 0 & 0 \\ -\varepsilon & \lambda_{cr} & 0 \\ \xi & \eta & \lambda_r \end{bmatrix} = \begin{bmatrix} 0 & 0 & 0 \\ -\varepsilon & 0 & 0 \\ \xi & \eta & 0 \end{bmatrix} + \begin{bmatrix} \lambda_{cr} & 0 & 0 \\ 0 & \lambda_{cr} & 0 \\ 0 & 0 & \lambda_r \end{bmatrix} \tag{42}$$

where $\phi = \beta - \alpha = R/2$, $\varepsilon = 2\alpha$, $\xi = \frac{\partial W}{\partial X}\big|_{\theta min}$, $\eta = \frac{\partial W}{\partial Y}\big|_{\theta min}$. Since the local rotational strength (magnitude) can be regarded as the scalar version of Rortex and the direction of the local rotational axis with the magnitude can be regarded as the vector version, Eq. (41) can be regarded as the tensor interpretation of Rortex which exactly represents the local rigidly rotational part of the velocity gradient tensor and consistent with the scalar and vector interpretations of Rortex. Eq. (42) contains the pure shearing and the stretching or compressing parts of the velocity gradient tensor. Because $\boldsymbol{S}$ has three real eigenvalues (multiple $\lambda_{cr}$ and $\lambda_r$), $\boldsymbol{S}$ itself implies no local rotation. Although the decomposition given by Eq. (40) is similar to Kolář's triple decomposition, Kolář's method is applied in the basic reference frame which remains unclear in 3D cases while our decomposition is a clear explicit expression which is obtained in a special coordinate frame determined by the orientation of the real eigenvector and the plane rotation given by Eq. (7). Additionally, Eq. (40) also sheds light on an analytical relation between Rortex and eigenvalues

$$\lambda_{ci} = \sqrt{\phi(\phi+\varepsilon)} \tag{43}$$

This expression will be applied in the following to examine the relations between Rortex and eigenvalue-based criteria.

## III. COMPARISON OF RORTEX AND EIGENVALUE-BASED VORTEX IDENTIFICATION CRITERIA

### A. Eigenvalue-based criteria

As earlier stated, most of the popular Eulerian vortex identification methods are based on the analysis of the velocity gradient tensor $\nabla \vec{v}$. More specifically, these methods are exclusively dependent on the eigenvalues of the velocity gradient tensor or the related invariants. Assume that $\lambda_1$, $\lambda_2$ and $\lambda_3$ are three eigenvalues. The characteristic equation can be written as

$$\lambda^3 + P\lambda^2 + Q\lambda + \tilde{R} = 0 \tag{44}$$

where

$$P = -(\lambda_1 + \lambda_2 + \lambda_3) = -tr(\nabla \vec{v}) \tag{45}$$

$$Q = \lambda_1\lambda_2 + \lambda_2\lambda_3 + \lambda_3\lambda_1 = -\frac{1}{2}(\text{tr}(\nabla\vec{v}^2) - \text{tr}(\nabla\vec{v})^2) \tag{46}$$

$$\tilde{R} = -\lambda_1\lambda_2\lambda_3 = -\det(\nabla\vec{v}) \tag{47}$$

$P$, $Q$ and $\tilde{R}$ are three invariants. For incompressible flow, according to continuous equation, we have $P = 0$.

Here we consider two representatives of eigenvalue-based criteria, namely the Q criterion and the $\lambda_{ci}$ criterion.

(1) Q criterion

The Q criterion is one of the most popular vortex identification method proposed by Hunt et al.[33] It identifies vortices of incompressible flow as fluid regions with positive second invariant, i.e. $Q > 0$. Meanwhile, a second condition requires the pressure in the vortical regions to be lower than the ambient pressure, despite often omitted in practice. Q is a measure of the vorticity magnitude in excess of the strain-rate magnitude, which can be expressed as

$$Q = \frac{1}{2}(\|\mathbf{\Omega}\|^2 - \|\mathbf{S}\|^2) \tag{48}$$

where $\mathbf{S}$ and $\mathbf{\Omega}$ are the symmetric and antisymmetric parts of the velocity gradient tensor, respectively

$$\mathbf{S} = \frac{1}{2}(\nabla\vec{v} + \nabla\vec{v}^{\text{T}}) = \begin{bmatrix} \frac{\partial u}{\partial x} & \frac{1}{2}\left(\frac{\partial u}{\partial y} + \frac{\partial v}{\partial x}\right) & \frac{1}{2}\left(\frac{\partial u}{\partial z} + \frac{\partial w}{\partial x}\right) \\ \frac{1}{2}\left(\frac{\partial v}{\partial x} + \frac{\partial u}{\partial y}\right) & \frac{\partial v}{\partial y} & \frac{1}{2}\left(\frac{\partial v}{\partial z} + \frac{\partial w}{\partial x}\right) \\ \frac{1}{2}\left(\frac{\partial w}{\partial x} + \frac{\partial u}{\partial z}\right) & \frac{1}{2}\left(\frac{\partial w}{\partial y} + \frac{\partial v}{\partial z}\right) & \frac{\partial w}{\partial z} \end{bmatrix} \tag{49}$$

$$\mathbf{\Omega} = \frac{1}{2}(\nabla \vec{v} - \nabla \vec{v}^{\mathrm{T}}) = \begin{bmatrix} 0 & \frac{1}{2}\left(\frac{\partial u}{\partial y} - \frac{\partial v}{\partial x}\right) & \frac{1}{2}\left(\frac{\partial u}{\partial z} - \frac{\partial w}{\partial x}\right) \\ \frac{1}{2}\left(\frac{\partial v}{\partial x} - \frac{\partial u}{\partial y}\right) & 0 & \frac{1}{2}\left(\frac{\partial v}{\partial z} - \frac{\partial w}{\partial x}\right) \\ \frac{1}{2}\left(\frac{\partial w}{\partial x} - \frac{\partial u}{\partial z}\right) & \frac{1}{2}\left(\frac{\partial w}{\partial y} - \frac{\partial v}{\partial z}\right) & 0 \end{bmatrix} \quad (50)$$

And $\|\cdot\|^2$ represents the Frobenius norm.

(2) $\lambda_{ci}$ criterion

The $\lambda_{ci}$ criterion is an extension of the $\Delta$ criterion and identical to the $\Delta$ criterion when zero threshold is applied.[29] When the velocity gradient tensor $\nabla \vec{v}$ has two complex eigenvalues, the local time-frozen streamlines exhibit a swirling flow pattern.[16] In this case, the eigen decomposition of $\nabla \vec{v}$ will give

$$\nabla \vec{v} = [\vec{v}_r \quad \vec{v}_{cr} \quad \vec{v}_{ci}] \begin{bmatrix} \lambda_r & 0 & 0 \\ 0 & \lambda_{cr} & \lambda_{ci} \\ 0 & -\lambda_{ci} & \lambda_{cr} \end{bmatrix} [\vec{v}_r \quad \vec{v}_{cr} \quad \vec{v}_{ci}]^{-1} \quad (51)$$

Here, $(\lambda_r, \vec{v}_r)$ is the real eigenpair and $(\lambda_{cr} \pm \lambda_{ci}, \vec{v}_{cr} \pm \vec{v}_{ci})$ the complex conjugate eigenpair. In the local curvilinear coordinate system $(c_1, c_2, c_3)$ spanned by the eigenvector $(\vec{v}_r, \vec{v}_{cr}, \vec{v}_{ci})$, the instantaneous streamlines are the same as pathlines and can be written as

$$c_1(t) = c_1(0)e^{\lambda_r t} \quad (52a)$$

$$c_2(t) = [c_2(0)\cos(\lambda_{ci}t) + c_3(0)\sin(\lambda_{ci}t)]e^{\lambda_{cr}t} \quad (52b)$$

$$c_3(t) = [c_3(0)\cos(\lambda_{ci}t) - c_2(0)\sin(\lambda_{ci}t)]e^{\lambda_{cr}t} \quad (52c)$$

where $t$ represents the time-like parameter and the constants $c_1(0)$, $c_2(0)$ and $c_3(0)$ are determined by the initial conditions. From Eq. (52b) and (52c), the period of orbit of a fluid particle is $2\pi/\lambda_{ci}$, so the imaginary part of the complex value $\lambda_{ci}$ is called swirling strength.

**B. Analytical relation and comparison between Rortex, Q criterion and $\lambda_{ci}$ criterion**

The analytical relation between $\phi$ and $\lambda_{ci}$ has been given by Eq. (43). The analytical relation between $\phi$ and Q can be obtained as

$$Q = \lambda_1 \lambda_2 + \lambda_2 \lambda_3 + \lambda_3 \lambda_1$$

$$= (\lambda_{cr} + i\lambda_{ci})(\lambda_{cr} - i\lambda_{ci}) + (\lambda_{cr} - i\lambda_{ci})\lambda_r + \lambda_r(\lambda_{cr} + i\lambda_{ci})$$

$$= \lambda_{cr}^2 + \lambda_{ci}^2 + 2\lambda_{cr}\lambda_r$$

$$= \lambda_{cr}^2 + \phi(\phi + \varepsilon) + 2\lambda_{cr}\lambda_r \tag{53}$$

Since $\lambda_{ci}$ and Q are eigenvalue-based, the same eigenvalues always yield the same values of $\lambda_{ci}$ and Q. In contrast, Rortex cannot be exclusively determined by eigenvalues. Assume that two velocity gradient tensors $\nabla\vec{v}|_A$ and $\nabla\vec{v}|_B$ have the same eigenvalues $\lambda_{cr} + i\lambda_{ci}$, $\lambda_{cr} - i\lambda_{ci}$ and $\lambda_r$ but different real eigenvectors. Through appropriate rotation matrices $Q|_A$ and $P|_A$, we can obtain $\nabla\vec{V}_{\theta min}|_A$ as

$$\nabla\vec{V}_{\theta min}\big|_A = P|_A Q|_A \nabla\vec{v}|_A (Q|_A)^T (P|_A)^T = \begin{bmatrix} \lambda_{cr} & \phi_A & 0 \\ -(\phi_A + \varepsilon_A) & \lambda_{cr} & 0 \\ \xi_A & \eta_A & \lambda_r \end{bmatrix} = \begin{bmatrix} 0 & \phi_A & 0 \\ -\phi_A & 0 & 0 \\ 0 & 0 & 0 \end{bmatrix} +$$

$$\begin{bmatrix} \lambda_{cr} & 0 & 0 \\ -\varepsilon_A & \lambda_{cr} & 0 \\ \xi_A & \eta_A & \lambda_r \end{bmatrix} \tag{54}$$

Similarly, through appropriate rotation matrices $Q|_B$ and $P|_B$, we have

$$\nabla\vec{V}_{\theta min}\big|_B = P|_B Q|_B \nabla\vec{v}|_B (Q|_B)^T (P|_B)^T = \begin{bmatrix} \lambda_{cr} & \phi_B & 0 \\ -(\phi_B + \varepsilon_B) & \lambda_{cr} & 0 \\ \xi_B & \eta_B & \lambda_r \end{bmatrix} = \begin{bmatrix} 0 & \phi_B & 0 \\ -\phi_B & 0 & 0 \\ 0 & 0 & 0 \end{bmatrix} +$$

$$\begin{bmatrix} \lambda_{cr} & 0 & 0 \\ -\varepsilon_B & \lambda_{cr} & 0 \\ \xi_B & \eta_B & \lambda_r \end{bmatrix} \tag{55}$$

Since the eigenvalues are identical, we have

$$Q|_A = Q|_B \tag{56}$$

$$\lambda_{ci}|_A = \lambda_{ci}|_B \tag{57}$$

and the following conditions

$$\phi_A(\phi_A + \varepsilon_A) = \lambda_{ci}^2 \tag{58}$$

$$\phi_B(\phi_B + \varepsilon_B) = \lambda_{ci}^2 \tag{59}$$

However, there is no further relation of $\phi_A$ and $\phi_B$, since four unknowns, i.e $\phi_A, \phi_B, \varepsilon_A, \varepsilon_B$ cannot be uniquely determined by two Eqs. (58) and (59). Therefore, in general, the rotational strength $\phi_A \neq \phi_B$. Consider a specific case. Two matrices

$$\nabla \vec{V}_{\theta min}\big|_A = \begin{bmatrix} 1 & 2 & 0 \\ -2 & 1 & 0 \\ \xi_A & \eta_A & 2 \end{bmatrix} = \begin{bmatrix} 0 & 2 & 0 \\ -2 & 0 & 0 \\ 0 & 0 & 0 \end{bmatrix} + \begin{bmatrix} 1 & 0 & 0 \\ 0 & 1 & 0 \\ \xi_A & \eta_A & 2 \end{bmatrix} \tag{60}$$

$$\nabla \vec{V}_{\theta min}\big|_B = \begin{bmatrix} 1 & 1 & 0 \\ -4 & 1 & 0 \\ \xi_B & \eta_B & 2 \end{bmatrix} = \begin{bmatrix} 0 & 1 & 0 \\ -1 & 0 & 0 \\ 0 & 0 & 0 \end{bmatrix} + \begin{bmatrix} 1 & 0 & 0 \\ -3 & 1 & 0 \\ \xi_B & \eta_B & 2 \end{bmatrix} \tag{61}$$

have the same eigenvalues $1 + 2i$, $1 - 2i$ and $2$. Certainly, we have $Q|_A = Q|_B = 9$ and $\lambda_{ci}|_A = \lambda_{ci}|_B = 2$. But the rotational strengths are quite different: $\phi_A = 2$ and $\phi_B = 1$.

From Eq. (43), we can find that the shearing effect $\varepsilon$ always exists in the imaginary part of the complex eigenvalues. Therefore, as long as eigenvalue-based criteria are dependent on the complex eigenvalues, they will be inevitably contaminated by shearing. Eqs. (43) and (53) indicate the shearing effect on $\lambda_{ci}$ and $Q$, respectively. The investigation of this contamination in some simple examples and realistic flows will be given in the following sections.

## IV. COMPARISON FOR SIMPLE EXAMPLES

### A. Rigid rotation

First, we consider 2D rigid rotation. The velocity in the polar coordinate system can be expressed as

$$\begin{cases} v_r = \omega r \\ v_\theta = 0 \end{cases} \tag{62}$$

Here, $\omega$ is a constant and represents the angular velocity. We assume $\omega > 0$, which means the flow field is rotating in clockwise order. Then, the velocity in the Cartesian coordinate system will be written as

$$\begin{cases} u = \omega y \\ v = -\omega x \end{cases} \tag{63}$$

In this simple case, we can analytically express Rortex, Q and $\lambda_{ci}$ as

$$R = 2\omega \tag{64}$$

$$Q = \omega^2 \tag{65}$$

$$\lambda_{ci} = \omega \tag{66}$$

It can be found that Rortex is exactly equal to vorticity.

Now consider the superposition of a prograde shearing motion, which is given by

$$\begin{cases} u = \sigma y, \ \sigma > 0 \\ v = 0 \end{cases} \tag{67}$$

$\sigma > 0$ implies that the shearing motion is consistent with the clockwise rigid rotation. The velocity becomes

$$\begin{cases} u = (\omega + \sigma)y \\ v = -\omega x \end{cases} \tag{68}$$

It can be easily verified that Eq. (68) fulfills 2D vorticity equations.

According to Eq. (40), the velocity gradient tensor is decomposed to

$$\begin{bmatrix} 0 & \omega + \sigma & 0 \\ -\omega & 0 & 0 \\ 0 & 0 & 0 \end{bmatrix} = \begin{bmatrix} 0 & \omega & 0 \\ -\omega & 0 & 0 \\ 0 & 0 & 0 \end{bmatrix} + \begin{bmatrix} 0 & \sigma & 0 \\ 0 & 0 & 0 \\ 0 & 0 & 0 \end{bmatrix} \tag{69}$$

which exactly presents the rigidly rotational part and the shearing part. The explicit expressions of Rortex, Q and $\lambda_{ci}$ are given by

$$R = 2\omega \tag{70}$$

$$Q = \omega(\omega + \sigma) \tag{71}$$

$$\lambda_{ci} = \sqrt{\omega(\omega + \sigma)} \tag{72}$$

It is expected that in this case the rigidly rotational part of fluids should not be affected by the shear motion. Only Rortex remains the same as no-shearing case and provides the precise rigidly rotational strength as expected, whereas Q and $\lambda_{ci}$ are altered by the shearing effect $\sigma$. Obviously, the stronger shearing will result in the larger alteration of Q and $\lambda_{ci}$, as shown in Fig. (1). In Fig. (1), the shearing effect $\sigma$ is normalized by the angular velocity $\omega$. With the increase of the shearing effect $\sigma$, Q and $\lambda_{ci}$ both indicate significant deviations from the values in the no-shearing case, which implies these criteria are prone to contamination by shearing and cannot reasonably represent the local rotation. In contrast, Rortex excludes the shearing effect and remains exactly twice of the angular velocity $\omega$ as the no-shearing case.

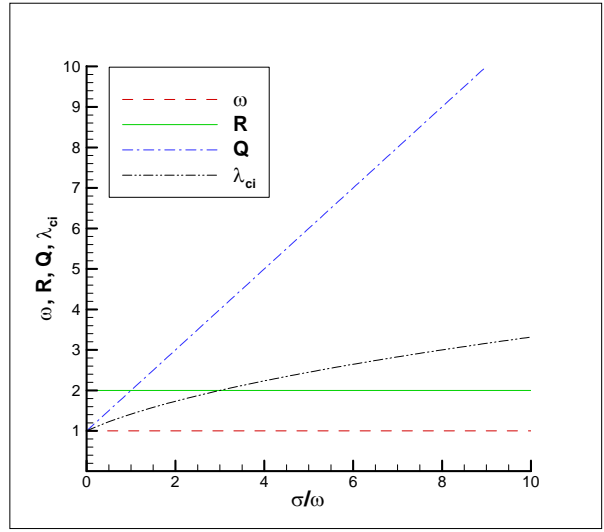

Fig. 1 R (Rortex), Q and $\lambda_{ci}$ as functions of $\sigma/\omega$ for 2D rigid rotation superposed by a prograde shearing motion.

**B. Burger vortex**

Here we examine the Burger vortex. This vortex has been widely used for modelling fine scales of turbulence. The Burger vortex is an exact steady solution of the Navier–Stokes equation, where the radial viscous diffusion of vorticity is dynamically balanced by vortex stretching due to an axisymmetric strain. The velocity components in the cylindrical coordinates for a Burger vortex can be written as

$$v_r = -\xi r \tag{73a}$$

$$v_\theta = \frac{\Gamma}{2\pi r}\left(1 - e^{\frac{-r^2 \xi}{2\nu}}\right) \tag{73b}$$

$$v_z = 2\xi z \tag{73c}$$

where $\Gamma$ is the circulation, $\xi$ the axisymmetric strain rate, and $\nu$ the kinematic viscosity. The Reynolds number for the vortex can be defined as $\text{Re} = \Gamma/(2\pi\nu)$. The velocity in the Cartesian coordinate system will be written as

$$u = -\xi x - \frac{\Gamma}{2\pi r^2}\left(1 - e^{\frac{-r^2 \xi}{2\nu}}\right) y \tag{74a}$$

$$v = -\xi y + \frac{\Gamma}{2\pi r^2}\left(1 - e^{\frac{-r^2 \xi}{2\nu}}\right) x \tag{74b}$$

$$w = 2\xi z \qquad (74c)$$

The analytical expressions of Rortex, Q and $\lambda_{ci}$ are given by

$$R = 2Re\xi\zeta \qquad (75)$$

$$Q = \xi^2[Re^2\zeta(\zeta + \varepsilon) - 3] \qquad (76)$$

$$\lambda_{ci} = Re\xi\sqrt{\zeta(\zeta + \varepsilon)} \qquad (77)$$

where $\tilde{r} = r\sqrt{\xi/\nu}$ and

$$\zeta = \frac{1}{\tilde{r}^2}\left[(1 + \tilde{r}^2)e^{-\frac{\tilde{r}^2}{2}} - 1\right]$$

$$\varepsilon = \frac{2}{\tilde{r}^2}\left[1 - \left(1 + \frac{\tilde{r}^2}{2}\right)e^{-\frac{\tilde{r}^2}{2}}\right]$$

Since Rortex and $\lambda_{ci}$ are equivalent to the Δ criterion with a zero threshold, the existence conditions of Rortex and $\lambda_{ci}$ are identical, namely, $\zeta > 0$, which yields a non-dimensional vortex size of $\tilde{r}_0 = 1.5852$, consistent with the result of Ref. 32.

Eqs. (76) and (77) indicate that the shearing part $\varepsilon$ will affect Q and $\lambda_{ci}$. To investigate this shearing effect, we consider the superposition of a shearing motion (with an appropriate external force term to fulfill the Navier-Stokes equations), which is given by

$$\begin{cases} u = -C\frac{Re\xi}{\tilde{r}_0^2}y \\ v = 0 \\ w = 0 \end{cases} \qquad (78)$$

where $C$ is a user-specified constant. We choose Re = 10 and $\xi = 1$. Figs. 2 to 4 demonstrate the iso-contours of Rortex, $\lambda_{ci}$ and Q in the xy plane for the Burgers vortex superposed with the shearing motion when $C$ is set to 1, 5 and 10, respectively. It can be obviously seen that the increase of the shearing motion slightly modifies the distribution of Rortex, but the rotational strength near the central part nearly remains constant. On the other hand, the distribution of $\lambda_{ci}$ and Q are significantly disturbed. The value of $\lambda_{ci}$ near the central part is increased from 6 to 14 and the value of Q near the center is increased from 40 to 180. This significant deviation demonstrates that these two criteria are prone to the contamination by shearing and the $\lambda_{ci}$ criterion is not a reliable measure of the local swirling strength at least when high shear strain exists.

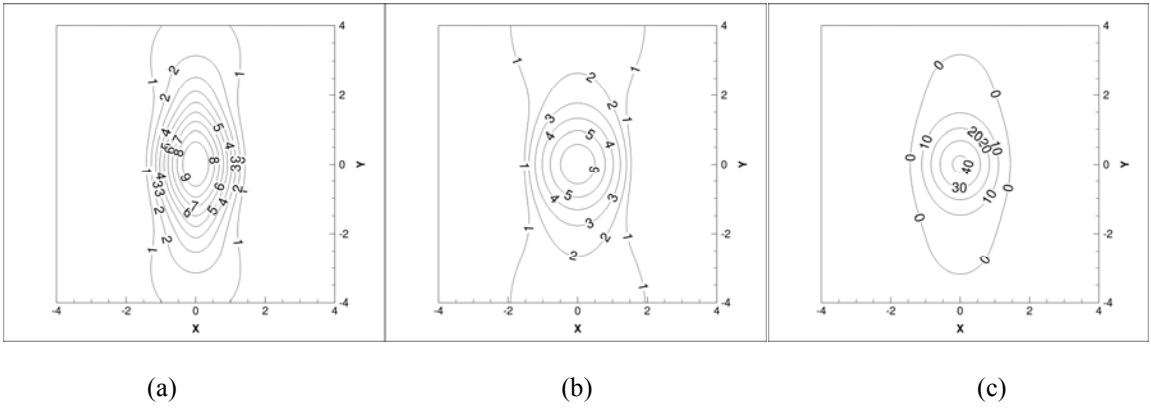

(a)　　　　　　　　　　　(b)　　　　　　　　　　　(c)

Fig. 2 Iso-contours of Rortex (a), $\lambda_{ci}$ (b) and Q (c) in the xy plane for the Burgers vortex superposed with the shearing motion ($C = 1$).

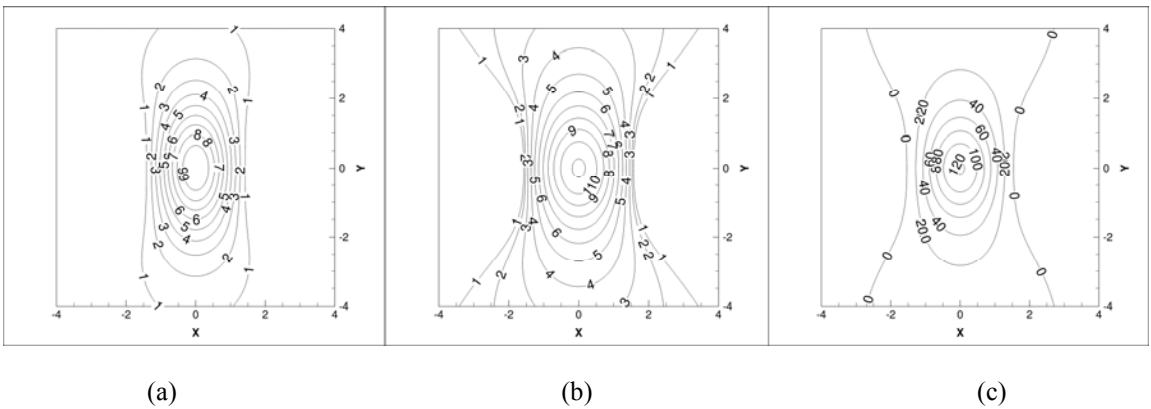

(a)　　　　　　　　　　　(b)　　　　　　　　　　　(c)

Fig. 3 Iso-contours of Rortex (a), $\lambda_{ci}$ (b) and Q (c) in the xy plane for the Burgers vortex superposed with the shearing motion ($C = 5$).

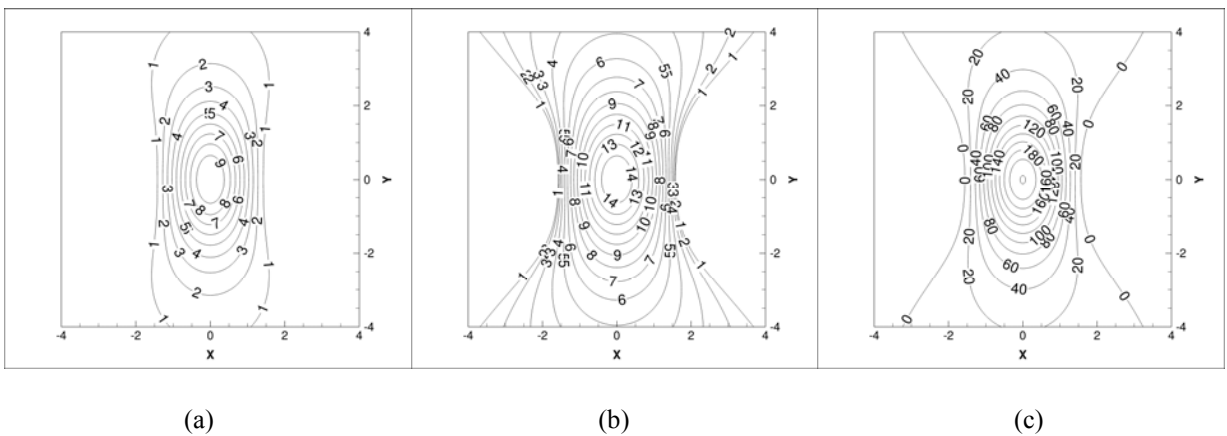

(a)　　　　　　　　　　　(b)　　　　　　　　　　　(c)

Fig. 4 Iso-contours of Rortex (a), $\lambda_{ci}$ (b) and Q (c) in the xy plane for the Burgers vortex superposed with the shearing motion ($C = 10$).

## C. Sullivan vortex

The Sullivan vortex is an exact solution to the Navier-Stokes equations for a three-dimensional axisymmetric two-celled vortex.[55] The two-celled vortex has an inner cell in which air flow descends from above and flows outward to meet a separate airflow that is converging radially. The mathematical form of the Sullivan Vortex is

$$v_r = -ar + \frac{6v}{r}\left(1 - e^{-\frac{ar^2}{2v}}\right) \tag{79a}$$

$$v_\theta = \frac{\Gamma}{2\pi r}\left(\frac{H\left(\frac{ar^2}{2v}\right)}{H(\infty)}\right) \tag{79b}$$

$$v_z = 2az\left(1 - 3e^{-\frac{ar^2}{2v}}\right) \tag{79c}$$

where

$$H(x) = \int_0^x e^{-t + 3\int_0^t [(1-e^{-\tau})/\tau]d\tau}\, dt \tag{80}$$

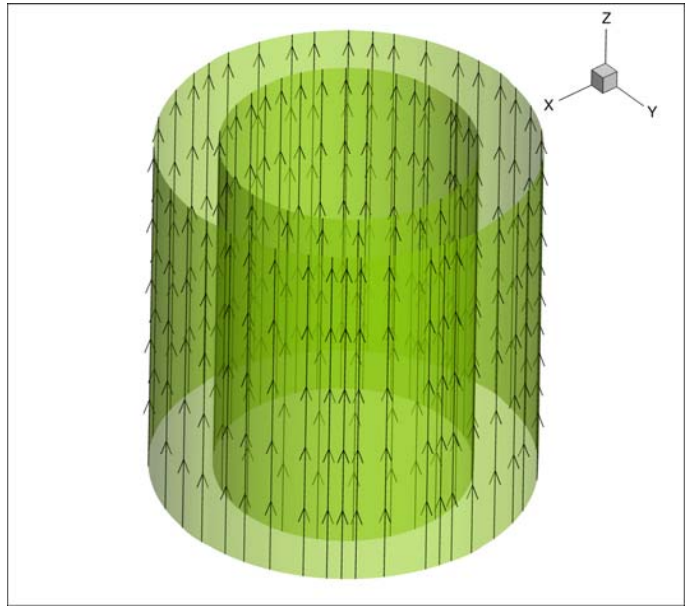

Fig. 5 Rortex vector lines for Sullivan vortex

In this case, we set $a = 1$, $\Gamma = 10$ and $v = 0.001$ to illustrate the local rotational axis. Fig. 5 shows the Rortex vector lines on the iso-surface which represent the local rotational axis. It can be seen that the local axis given by Rortex is consistent with the global rotation axis, that is, the z axis, which means the direction of Rortex is physically reasonable.

## V. COMPARISON FOR REALISTIC FLOWS

Here we use the DNS data of late boundary layer transition on a flat plate to compare Rortex with Q and $\lambda_{ci}$. The DNS data are generated by a DNS code called DNSUTA.[9] A sixth-order compact scheme is applied in the streamwise and normal directions. In the spanwise direction where periodic conditions are applied, the pseudo-spectral method is used. In order to eliminate the spurious numerical oscillations caused by central difference schemes, an implicit sixth-order compact filter is applied to the primitive variables after a specified number of time steps. The simulation was performed with near 60 million grid points and over 400,000 time steps at a free stream Mach number of 0.5. For the detailed case setup, please refer to Ref. 9.

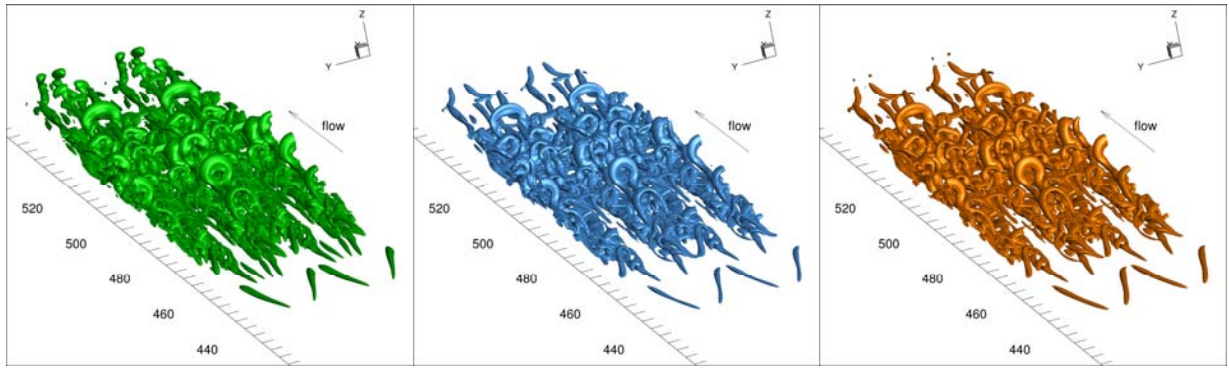

Fig. 6 Isosufaces of Rortex, $\lambda_{ci}$ and Q for late boundary layer transition

Although all methods illustrate the similar iso-surfaces of vortical structures as shown in Fig. 6, the values of $\lambda_{ci}$ and Q can be found contaminated by shearing. Examine three points A, B and C on the Rortex and $\lambda_{ci}$ iso-surfaces of the quasi-streamwise vortical structure as shown in Fig. 7. Point A is located on both the Rortex and $\lambda_{ci}$ iso-surfaces, B on the Rortex iso-surface only and C on the $\lambda_{ci}$ iso-surface only. The corresponding velocity gradient tensors of A, B and C are given by Eq. (81). The eigenvalues, the magnitudes of Rortex and shearing components are provided in Table 2. From Table 2, we can find that A and B possess the same local rotational strength with different eigenvalues, while A and C have the same imaginary value of the complex eigenvalues but different local rotational strength. The shearing parts are so strong that the $\lambda_{ci}$ criterion will be seriously contaminated. Especially for point C, the shearing component $\varepsilon = 0.81$ is significantly larger than the actual local rotation strength R = 0.018, making point C being mistaken for a point with large swirling strength by the $\lambda_{ci}$ criterion. From Figure 7, it can be found that point B which has a strong rotation (B=0.06) is missed by the $\lambda_{ci}$ criterion, but point C which contains a weak rotation (R=0.018) is mis-identified by the $\lambda_{ci}$ criterion. The Q criterion as shown in Fig. 8 will indicate a similar result of contamination, so the detailed analysis is omitted here.

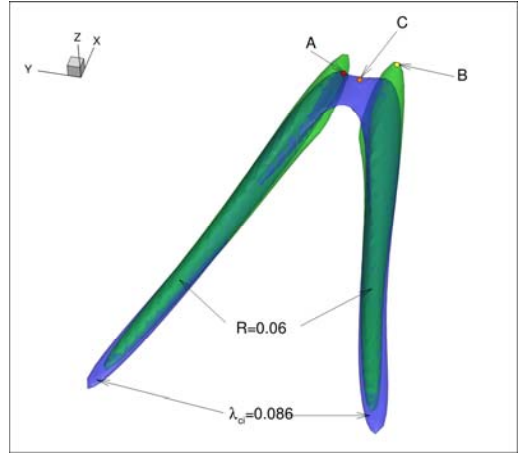

Fig. 7 Iso-surfaces of Rortex and $\lambda_{ci}$

$$\nabla\vec{v}|_A = \begin{bmatrix} 0.0622 & 0.215 & 0.560 \\ -0.00733 & 0.00312 & 0.0562 \\ -0.0121 & -0.0514 & -0.0664 \end{bmatrix} \tag{81a}$$

$$\nabla\vec{v}|_B = \begin{bmatrix} 0.0498 & 0.211 & 0.361 \\ -0.00109 & -0.00986 & 0.0697 \\ -0.00625 & -0.0417 & -0.0406 \end{bmatrix} \tag{81b}$$

$$\nabla\vec{v}|_C = \begin{bmatrix} 0.0372 & -0.000897 & 0.818 \\ 0.000064 & 0.0281 & 0.00012 \\ -0.0124 & 0.000250 & -0.0674 \end{bmatrix} \tag{81c}$$

Table 2. Eigenvalues ($\lambda_r$, $\lambda_{cr}$ and $\lambda_{ci}$), Rortex strengths (R) and shearing components ($\varepsilon$) of point A, B and C

|  | A | B | C |
| --- | --- | --- | --- |
| $\lambda_r$ | 0.0197 | 0.0162 | 0.0281 |
| $\lambda_{cr}$ | -0.00104 | -0.00843 | -0.0151 |
| $\lambda_{ci}$ | 0.086 | 0.059 | 0.086 |
| R | 0.06 | 0.06 | 0.018 |
| $\varepsilon$ | 0.218 | 0.0866 | 0.81 |

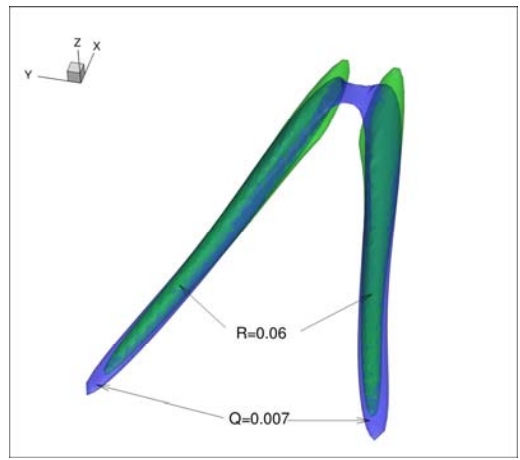

Fig. 8 Iso-surfaces of Rortex and Q

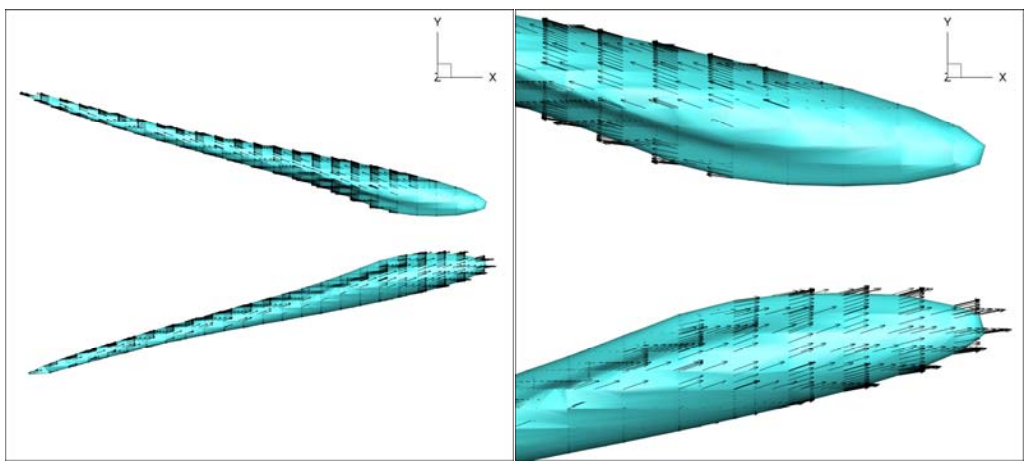

Fig. 9 Rortex vector on the leg part of the vorical structure

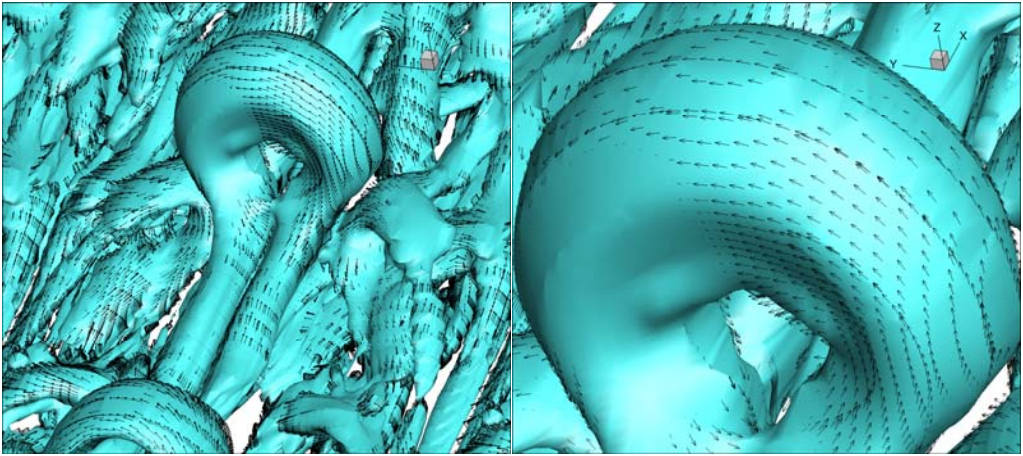

Fig. 10 Rortex vector on the vortex ring

Since Rortex is a vector quantity, we can visualize the local rotation axis of the vortex structures. Figs. 9 and 10 demonstrate that the Rortex vector is actually tangent to the iso-surface of Rortex. Assume a point $P$ is located on the iso-surface and a point $P^*$ is on the direction of Rortex vector at $P$, as shown in Fig. 11. According to Definition 1, when $P^*$ limits toward $P$, only the velocity along the local rotation axis Z can change. Correspondingly, only the component along the local rotation axis Z of the velocity gradient tensor can change. So, the component of the velocity gradient tensor in the XY plane will not change, which means $P^*$ will be located on the same iso-surface in the limit and Rortex vector is tangent to the iso-surface of Rortex at point $P$.

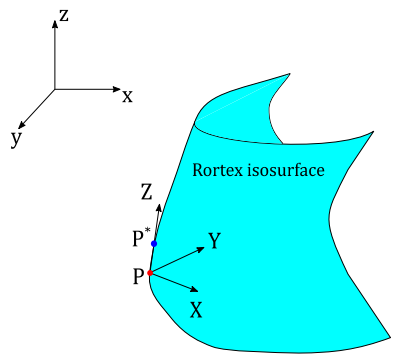

Fig. 11 Illustration of Rortex vector at point $P$.

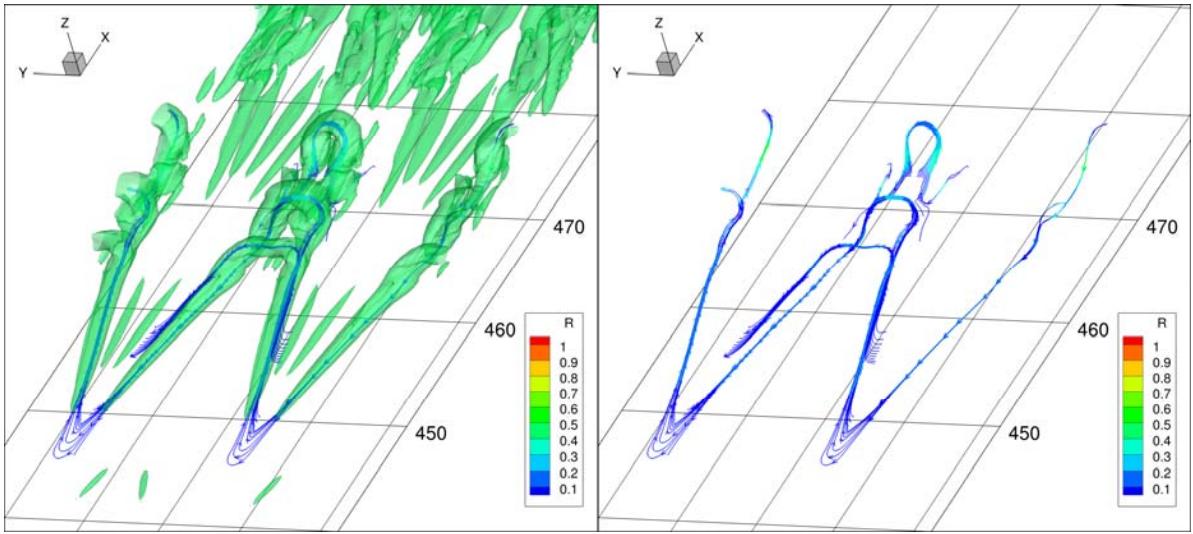

Fig. 12 Rortex lines for hairpin vortex

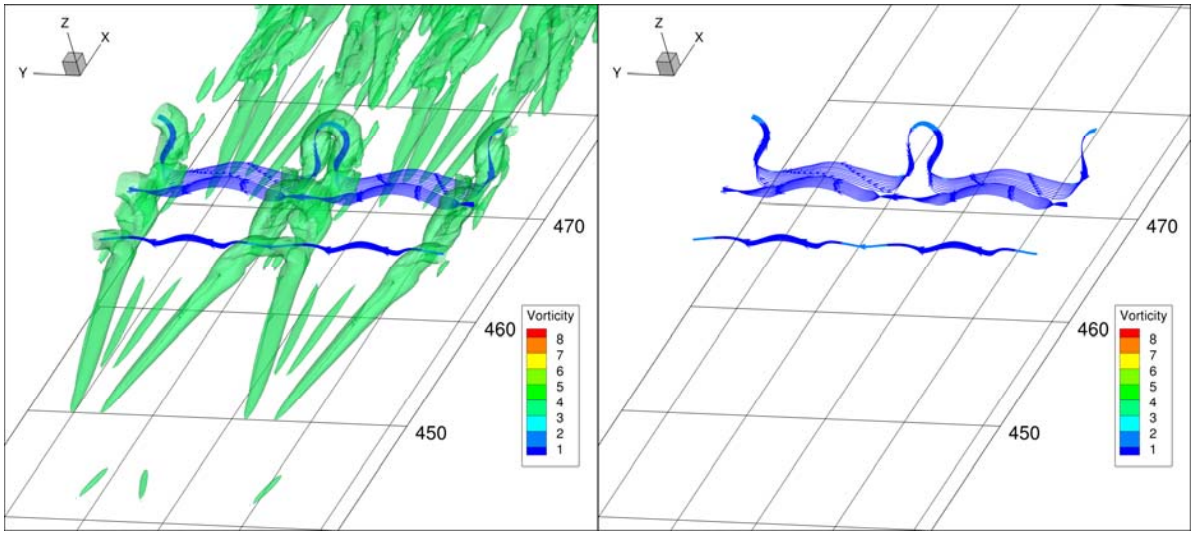

Fig. 13 Vorticity lines for hairpin vortex

Fig. 12 shows the structures of Rortex lines and Fig. 13 demonstrates vorticity lines. Both pass the same seed points. As can be seen, vorticity lines can only represent the ring part of the hairpin vortex. In contrast, Rortex lines can provide a skeleton of the whole hairpin vortex. It is expected that Rortex lines will offer a new perspective to analyze the vortical structures.

## VI. CONCLUSIONS

In the present study, an alternative eigenvector-based definition of Rortex is introduced. A systematic interpretation of scalar, vector and tensor versions of Rortex is presented to provide a unified characterization of the local fluid rotation. Several conclusions are summarized as follows:

(1) The real eigenvector of the velocity gradient tensor is used to determine the direction of Rortex, which represents the possible axis of the local fluid rotation, and the rotational strength obtained in the plane perpendicular to the possible local axis is defined as the magnitude of Rortex.

(2) Eigenvalue-based criteria are exclusively determined by the eigenvalues of the velocity gradient tensor. If two points have the same eigenvalues, they are located on the same iso-surface. But Rortex cannot be exclusively determined by the eigenvalues. Even if two points have the same eigenvalues, the magnitudes of Rortex are generally different.

(3) The existing eigenvalue-based methods can be seriously contaminated by shearing. Since shearing always manifests its effect on the imaginary part of the complex eigenvalues, any criterion associated with the complex eigenvalues will be prone to contamination by shear. While Rortex eliminates the contamination and thus can accurately quantify the local rotational strength.

(4) Rortex can identify the local rotational axis and provide the precise local rotational strength, thereby can reasonably represent the local rigidly rotation of fluids.

(5) In contrast to eigenvalue-based criteria, as a vector quantity, not only the iso-surface of Rortex but also Rortex vector field and Rortex lines can be used to visualize and investigate vortical structures.

(6) A new velocity gradient tensor decomposition is proposed. The velocity gradient tensor is decomposed to a rigid rotation part and a non-rotational part including shearing, stretching and compression, different from the traditional symmetric and anti-symmetric tensor decomposition.

(7) Since both local rotation axis and magnitude of Rortex is uniquely determined by the velocity gradient tensor without any dynamics involved, Rortex is a mathematical definition of fluid kinematics.

(8) Our new implementation to calculate Rortex dramatically improves the computational efficiency. The calculation time of the present method is reduced by one order of magnitude compared to our previous methods and comparable to that of the $\lambda_{ci}$ criterion.

**ACKNOWLEDGEMENTS**

This work was supported by the Department of Mathematics at University of Texas at Arlington and AFOSR grant MURI FA9559-16-1-0364. The authors are grateful to Texas Advanced Computing Center (TACC) for providing computation hours. This work is accomplished by using Code DNSUTA which was released by Dr. Chaoqun Liu at University of Texas at Arlington in 2009. The name of Rortex is credited to the discussion with many colleagues in the WeChat groups.

**Appendix A**

In Appendix A, an analytical solution for the eigenvalues of the velocity gradient tensor is presented. Let **A** be a matrix representation of the velocity gradient tensor in the original $xyz$-frame

$$\mathbf{A} = \begin{bmatrix} \frac{\partial u}{\partial x} & \frac{\partial u}{\partial y} & \frac{\partial u}{\partial z} \\ \frac{\partial v}{\partial x} & \frac{\partial v}{\partial y} & \frac{\partial v}{\partial z} \\ \frac{\partial w}{\partial x} & \frac{\partial w}{\partial y} & \frac{\partial w}{\partial z} \end{bmatrix} \tag{A1}$$

and $\lambda$ the eigenvalue. The characteristic equation of the matrix **A** is given by

$$\lambda^3 + P\lambda^2 + Q\lambda + \tilde{R} = 0 \tag{A2}$$

where

$$P = -(\lambda_1 + \lambda_2 + \lambda_3) = -\text{tr}(\mathbf{A}) \tag{A3}$$

$$Q = \lambda_1\lambda_2 + \lambda_2\lambda_3 + \lambda_3\lambda_1 = -\frac{1}{2}(\text{tr}(\mathbf{A}^2) - \text{tr}(\mathbf{A})^2) \tag{A4}$$

$$\tilde{R} = -\lambda_1\lambda_2\lambda_3 = -\det(\mathbf{A}) \tag{A5}$$

Here, tr represents the trace of the matrix and det the determinant. The cubic equation (A2) can be solved by a robust algorithm to minimize roundoff error.[56] Here we are only concerned about the case of the existence of two complex roots as the existence of three real roots imply no local rotation. First, we compute

$$S \equiv \frac{P^2 - 3Q}{9} \tag{A6}$$

$$T \equiv \frac{2P^3 - 9PQ + 27\tilde{R}}{54} \tag{A7}$$

If $T^2 > S^3$, the cubic equation has two complex roots. By computing

$$A = -\text{sgn}(T)\left[|T| + \sqrt{T^2 - S^3}\right]^{1/3} \tag{A8}$$

$$B = \begin{cases} S/A & (A = 0) \\ 0 & (A = 0) \end{cases} \tag{A9}$$

where sgn is the sign function, the three roots can be written as

$$\lambda_1 = -\frac{1}{2}(A + B) - \frac{P}{3} + i\frac{\sqrt{3}}{2}(A - B) \tag{A10}$$

$$\lambda_2 = -\frac{1}{2}(A + B) - \frac{P}{3} - i\frac{\sqrt{3}}{2}(A - B) \tag{A11}$$

$$\lambda_3 = (A + B) - \frac{P}{3} \tag{A12}$$

Because A and B are both real, $\lambda_1$ and $\lambda_2$ are the complex eigenvalues and $\lambda_3$ is the real eigenvalue.

**Appendix B**

Here, we derive the analytical expression of the normalized real eigenvector $\vec{r}$ corresponding to the real eigenvalue $\lambda_r$. Also, we focus on the case of the existence of two complex eigenvalues and one real eigenvalue. In this case, the normalized real eigenvector is unique (up to sign). Assuming that **A** is a matrix representation of the velocity gradient tensor and $\vec{r}^* = [r_x^*, r_y^*, r_z^*]^T$ represents an unnormalized eigenvector corresponding to $\lambda_r$, we can obtain the following equation

$$\mathbf{A}\vec{r}^* = \lambda_r \vec{r}^* \tag{B1}$$

Eq. (B1) can be rewritten as

$$\begin{bmatrix} \frac{\partial u}{\partial x} - \lambda_r & \frac{\partial u}{\partial y} & \frac{\partial u}{\partial z} \\ \frac{\partial v}{\partial x} & \frac{\partial v}{\partial y} - \lambda_r & \frac{\partial v}{\partial z} \\ \frac{\partial w}{\partial x} & \frac{\partial w}{\partial y} & \frac{\partial w}{\partial z} - \lambda_r \end{bmatrix} \begin{bmatrix} r_x^* \\ r_y^* \\ r_z^* \end{bmatrix} = 0 \tag{B2}$$

By checking three first minors

$$\Delta_x = \begin{vmatrix} \frac{\partial v}{\partial y} - \lambda_r & \frac{\partial v}{\partial z} \\ \frac{\partial w}{\partial y} & \frac{\partial w}{\partial z} - \lambda_r \end{vmatrix} \tag{B3}$$

$$\Delta_y = \begin{vmatrix} \frac{\partial u}{\partial x} - \lambda_r & \frac{\partial u}{\partial z} \\ \frac{\partial w}{\partial x} & \frac{\partial w}{\partial z} - \lambda_r \end{vmatrix} \tag{B4}$$

$$\Delta_z = \begin{vmatrix} \frac{\partial u}{\partial x} - \lambda_r & \frac{\partial u}{\partial y} \\ \frac{\partial v}{\partial x} & \frac{\partial v}{\partial y} - \lambda_r \end{vmatrix} \tag{B5}$$

we can find the maximum absolute value

$$\Delta_{max} = \max(|\Delta_x|, |\Delta_y|, |\Delta_z|) \tag{B6}$$

(Note: not all the minors will be equal to zero, thus $\Delta_{max} > 0$. Otherwise, we will arrive at a contradiction that the normalized real eigenvector is nonunique, or the real eigenvector is a zero vector.)

If $\Delta_{max} = |\Delta_x|$, we can set

$$r_x^* = 1 \tag{B7}$$

By solving

$$\begin{bmatrix} \frac{\partial v}{\partial y} - \lambda_r & \frac{\partial v}{\partial z} \\ \frac{\partial w}{\partial y} & \frac{\partial w}{\partial z} - \lambda_r \end{bmatrix} \begin{bmatrix} r_y^* \\ r_z^* \end{bmatrix} = \begin{bmatrix} -\frac{\partial v}{\partial x} \\ -\frac{\partial w}{\partial x} \end{bmatrix} \tag{B8}$$

we obtain the other two components of $\vec{r}^*$ as

$$r_y^* = \frac{-\left(\frac{\partial w}{\partial z} - \lambda_r\right)\frac{\partial v}{\partial x} + \frac{\partial v \partial w}{\partial z \partial x}}{\left(\frac{\partial v}{\partial y} - \lambda_r\right)\left(\frac{\partial w}{\partial z} - \lambda_r\right) - \frac{\partial v \partial w}{\partial z \partial y}} \tag{B9}$$

$$r_z^* = \frac{\frac{\partial w \partial v}{\partial y \partial x} - \left(\frac{\partial v}{\partial y} - \lambda_r\right)\frac{\partial w}{\partial x}}{\left(\frac{\partial v}{\partial y} - \lambda_r\right)\left(\frac{\partial w}{\partial z} - \lambda_r\right) - \frac{\partial v \partial w}{\partial z \partial y}} \tag{B10}$$

Similarly, if $\Delta_{max} = |\Delta_y|$, we chose

$$r_y^* = 1 \tag{B11}$$

By solving

$$\begin{bmatrix} \frac{\partial u}{\partial x} - \lambda_r & \frac{\partial u}{\partial z} \\ \frac{\partial w}{\partial x} & \frac{\partial w}{\partial z} - \lambda_r \end{bmatrix} \begin{bmatrix} r_x^* \\ r_z^* \end{bmatrix} = \begin{bmatrix} -\frac{\partial u}{\partial y} \\ -\frac{\partial w}{\partial y} \end{bmatrix} \tag{B12}$$

we have

$$r_x^* = \frac{-\left(\frac{\partial w}{\partial z} - \lambda_r\right)\frac{\partial u}{\partial y} + \frac{\partial u \partial w}{\partial z \partial y}}{\left(\frac{\partial u}{\partial x} - \lambda_r\right)\left(\frac{\partial w}{\partial z} - \lambda_r\right) - \frac{\partial u \partial w}{\partial z \partial x}} \tag{B13}$$

$$r_z^* = \frac{\frac{\partial w}{\partial x}\frac{\partial u}{\partial y}-\left(\frac{\partial u}{\partial x}-\lambda_r\right)\frac{\partial w}{\partial y}}{\left(\frac{\partial u}{\partial x}-\lambda_r\right)\left(\frac{\partial w}{\partial z}-\lambda_r\right)-\frac{\partial u}{\partial z}\frac{\partial w}{\partial x}} \tag{B14}$$

In the case of $\Delta_{max}= |\Delta_z|$, we set

$$r_z^* = 1 \tag{B15}$$

By solving

$$\begin{bmatrix} \frac{\partial u}{\partial x} - \lambda_r & \frac{\partial u}{\partial y} \\ \frac{\partial v}{\partial x} & \frac{\partial v}{\partial y} - \lambda_r \end{bmatrix} \begin{bmatrix} r_x^* \\ r_y^* \end{bmatrix} = \begin{bmatrix} -\frac{\partial u}{\partial z} \\ -\frac{\partial v}{\partial z} \end{bmatrix} \tag{B16}$$

we can find the other two components of $\vec{r}^*$ as

$$r_x^* = \frac{-\left(\frac{\partial v}{\partial y}-\lambda_r\right)\frac{\partial u}{\partial z}+\frac{\partial u}{\partial y}\frac{\partial v}{\partial z}}{\left(\frac{\partial u}{\partial x}-\lambda_r\right)\left(\frac{\partial v}{\partial y}-\lambda_r\right)-\frac{\partial u}{\partial y}\frac{\partial v}{\partial x}} \tag{B17}$$

$$r_y^* = \frac{\frac{\partial v}{\partial x}\frac{\partial u}{\partial z}-\left(\frac{\partial u}{\partial x}-\lambda_r\right)\frac{\partial v}{\partial z}}{\left(\frac{\partial u}{\partial x}-\lambda_r\right)\left(\frac{\partial v}{\partial y}-\lambda_r\right)-\frac{\partial u}{\partial y}\frac{\partial v}{\partial x}} \tag{B18}$$

And the normalized real eigenvector $\vec{r}$ will be

$$\vec{r} = \vec{r}^*/|\vec{r}^*| \tag{B19}$$